\documentclass[preprint2]{aastex}

\shorttitle{Quiet Photosphere Horizontal Magnetic Fields}
\shortauthors{Harvey et al.}

\begin{document}

\title{Seething Horizontal Magnetic Fields in the Quiet Solar Photosphere}

\author{J. W. Harvey, D. Branston, C. J. Henney, C. U. Keller\altaffilmark{1},
and the SOLIS and GONG Teams} 
\affil{National Solar Observatory, Tucson, AZ 85726}
\altaffiltext{1}{Sterrekundig Instituut, Utrecht University, NL-3508 TA
Utrecht, The Netherlands}

\begin{abstract}
The photospheric magnetic field outside of active regions and the
network has a ubiquitous and dynamic line-of-sight component that
strengthens from disk center to limb as expected for a nearly horizontal
orientation. This component shows a striking time variation with
an average temporal rms near the limb of 1.7 G at $\sim$3\arcsec\ resolution.
In our moderate resolution observations the nearly horizontal component has
a frequency variation power law exponent of -1.4 below 1.5 mHz and is
spatially patchy on scales up to $\sim$15 arcsec. The field may be
a manifestation of changing magnetic connections between eruptions
and evolution of small magnetic flux elements in response to convective
motions. It shows no detectable latitude or longitude variations.
\end{abstract}

\keywords{Sun: magnetic fields --- Sun: photosphere}

\section{Introduction}

It has been known for more than thirty years that the `quiet' photosphere
contains magnetic fields \citep{liv71}. Most prominent are compact,
mixed-polarity flux elements of the order of 10$^{16}$ Mx and arcsec
sizes that vary over tens of minutes \citep{liv75,smi75}. The
quiet-Sun magnetic field is known by various names with the term
internetwork (IN) field in wide use. Small fluxes and sizes with
rapid time changes make the compact IN fields difficult to observe and
characterize (e.g. Keller et al. 1994). \citet{lin99} detected an 
additional component consisting of a close association between
granulation and a fluctuating $\sim$1 G vertical component of the quiet
Sun magnetic field that they call a granular magnetic field.

Most previous IN field observations have been made at or near disk center
using line-of-sight (LOS) magnetograms that reveal properties of
the vertical component of the IN field (e.g.
Socas-Navarro, Mart\'{\i}nez Pillet, \& Lites 2004).
There is scant information about the center-to-limb
variation of the IN or its possible horizontal components. \citet{mar88} 
presented IN field observations showing little, if any, center-to-limb
variation of the LOS component implying that the IN
fields are more isotropically oriented than the network fields. \citet{lit96}
used vector magnetograms near disk center to discover
sporadic short-lived (5 min), arcsec-scale horizontal IN field
elements that they named HIFs. They associated HIFs with the eruption
of small bipolar elements of magnetic flux from the solar interior.
\citet{meu98} made sensitive 1-D scans across the disk and concluded
that the IN field consisted of relatively strong, mainly vertically-oriented
features and weaker, mainly horizontal components.
\citet{dep02}, observing at a heliocentric angle of 38\arcdeg\, found several
LOS magnetic features, longer lived than HIFs, which were 
interpreted as being mainly horizontally oriented and closely related to
granular flow dynamics. 

In this Letter we present results from time sequences of LOS component
magnetograms of the quiet Sun made with the SOLIS vector spectromagnetograph
(VSM) \citep{kel03} and the GONG network instruments \citep{har88}.
Using different methods, the instruments provide the difference between
the wavelengths of Zeeman sensitive lines in right and left circularly
polarized light. These differences are expressed as the homogeneous LOS
field strength in gauss that would produce the measured splitting.
Since the field is generally inhomogeneous, we obtain only lower limits
of true LOS field strengths. Our observations are unique in combining full-disk
coverage and comparatively high time cadence, with good sensitivity and
moderate spatial resolution. These properties have revealed that
there is a ubiquitous, spatially-structured, nearly horizontal field
component that varies strikingly over a wide range of time periods. 

\section{Observations and Results}

We made time sequences of LOS magnetograms with the VSM (90 s cadence,
3.2 h duration on 2006 December 15, 1\farcs1 pixels) and with GONG 
(10 min cadence of 10 min averages coinciding with the VSM data and
also a 7 h duration on 2006 December 6, 2\farcs5 pixels).
Each time sequence was registered to a
fixed solar image centering and then disk features were rotated to a
selected time using an assumed representation of solar rotation. These
steps led to movies that emphasized real solar changes.

Near disk center the movies show relatively slow variations consisting
of evolution of the network fields and the mixed-polarity IN fields.
Increasingly obvious away from disk center, another mixed-polarity, more 
dynamic component is present everywhere in otherwise quiet areas. It is
patchy with sizes ranging from our resolution limit of a few arcsec to
$\sim$15\arcsec\ that remain visible for a few minutes to more than 15 min
at our noise levels of $\sim$1 and $\sim$0.2 G per pixel for GONG and VSM
data respectively.

\begin{figure}[!h]
\includegraphics[width=3.25in,keepaspectratio=true,clip=true]{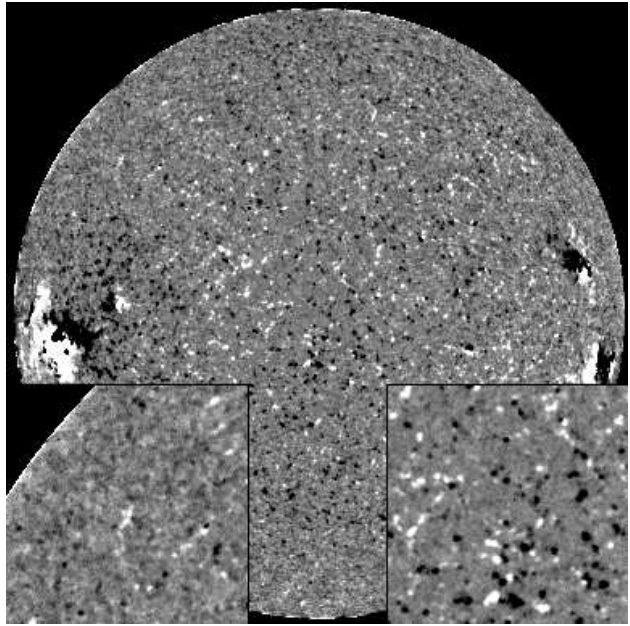}
\caption{Line-of-sight component of the photospheric magnetic field
averaged from 18:42--18:49 UT on 2006 December 6 observed with the GONG
instrument at Big Bear Solar Observatory. {\it Insets} are twice-magnified
pieces from the upper left ({\it left}) and disk center ({\it right}).
Note increased background mottling toward the limb compared to near
disk center. The display saturates at $\pm$15 G.}
\end{figure}

The heretofore unrecognized patchy field is visible in single observations
such as Figure 1 as an increasingly mottled background structure as one
looks from disk center to the limb. In contrast, the visibility of
network fields decreases toward the limb. Figure 2 better
emphasizes these dynamic background structures by subtracting a long time 
average from a single magnetogram. In addition, we prepared Figure 3
from 7 h of magnetograms to study the center-to-limb behavior of the changing
fields. It shows the temporal rms of the time-varying fields over the solar
disk. To emphasize the patchy field variations, we excluded
any measurements with absolute field strengths $>$5.5 G and set 
such areas to white in the figure. Remaining is the temporal rms variation of
the dynamic background field and other sources of changes. These other
sources include seeing and registration variations, proper motion of
magnetic features, appearance, disappearance, and shape changes of
magnetic features, barely resolved IN fields, and instrumental 
noise. The latter is a combination of camera and photon noise---being
relatively low near the bright disk center and greater near the darker limb.
We modeled this
noise and find it to vary with radius only slowly except close to the limb.
The majority of the signal variation in Figure 3 is caused by the dynamic
background field.

\begin{figure}[!h]
\includegraphics[width=3.25in,keepaspectratio=true,clip=true]{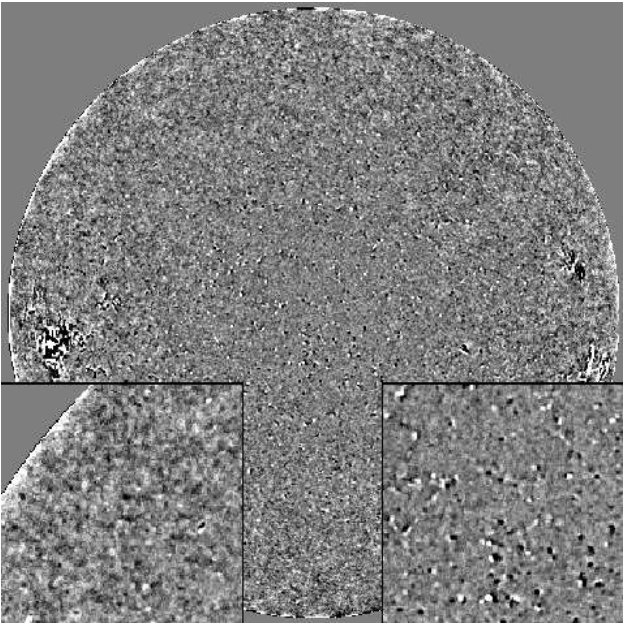}
\caption{Difference between the data of Figure 1 and a 16:42--23:34 UT
averaged magnetogram to emphasize the increased background structure near the
limb compared to near disk center. Same format as Figure 1. Display
saturates at $\pm$7.5 G.}
\end{figure}

\begin{figure}[!h]
\includegraphics[width=3.25in,keepaspectratio=true,clip=true]{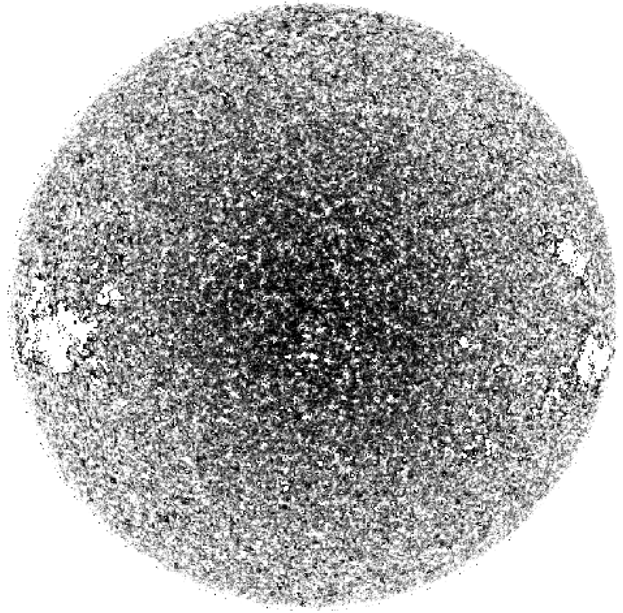}
\caption{Temporal rms variation of the LOS magnetic field from 16:42--23:34 UT.
Black corresponds to 1.3 G and white to 2.3 G. Regions with measured absolute
LOS field strengths $>$5.5 G were not included in the calculation of the
temporal rms and are set to white. Note the increase of the rms toward
the limb and the absence of any large-scale departure from radial symmetry.}
\end{figure}

The obvious increase of the magnetic field fluctuations toward the limb
suggests that the dynamic features are mainly horizontally oriented.
Figure 4 (upper curve) is a radial average of the LOS data in Figure 3
(excluding strong fields) and supports this idea. Horizontally-oriented
structures should strengthen with the sine of heliocentric angle, i.e.,
a linear increase with distance from disk center. The upper curve nicely
shows a linear trend but also contains instrumental noise (and a seeing-noise
spike at the limb). The dashed curve is a model of camera and
photon noise. The lower curve is the quadratic difference of the
observed rms minus the instrumental noise model.
The corrected rms is not linear, suggesting that the
fluctuating field is not strictly horizontal. The dotted curve is a
model for which the field is inclined to the vertical by 74\arcdeg\ and
fits the data.
The rms near the limb could also be reduced by loss of resolution due to
foreshortening and a possible height variation of the dynamic field, factors
that would make the inferred field direction more nearly horizontal.

\begin{figure}[!h]
\includegraphics[width=3.0in,keepaspectratio=true,clip=true]{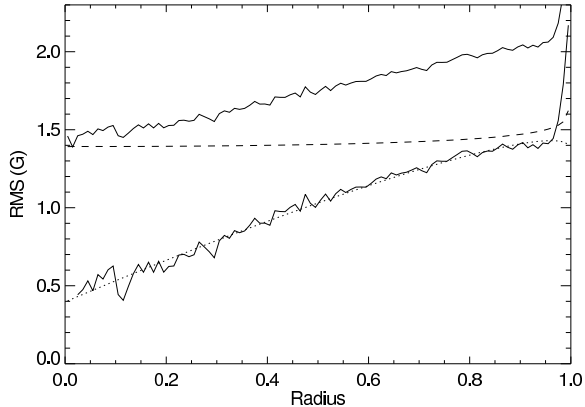}
\caption{({\it upper curve}) Radial average of the data in Figure 3 excluding
regions with measured absolute fields $>$5.5 G. ({\it dashed curve})
Model of noise due to the camera and photon statistics. ({\it lower
curve}) Observed rms corrected for the noise model. ({\it dotted curve}) Model
of the corrected rms.}
\end{figure}

VSM data confirm the GONG results with lower noise and higher spatial
resolution and cadence. Figure 5{\it a} is a 1880\arcsec\ by 83\arcsec\
cut from one frame of a VSM time series of LOS magnetograms (offset
$\sim$2\arcmin\ from disk center). The time variation
over 3.2 h along a trace through this area is shown in Figure 5{\it b}.
The varying horizontal magnetic field is seen near the edges as mottling.
In comparison, the mottling is nearly absent in quiet areas near disk center.
The temporal rms of network-free regions near disk center measures 0.8 G
while near the limb it is 1.9 G.
Quadratically subtracting these values gives 1.7 G as the temporal rms
of the horizontal field near the limb. This can be compared with the
lower resolution GONG value of $\sim$1.4 G near the limb (bottom curve
in Figure 4). Higher spatial resolution data would no doubt give a larger
value for the temporal rms of the horizontal field. Figure 5{\it c} is data 
from the GONG instrument located at Cerro Tololo processed to try to
match the VSM data. The GONG data is noisier and the registration
is imperfect, especially just left of center, but similarity of
the background solar signals is evident here and in movies.
Figure 5{\it d} is a series of power 
spectra of the data in Figure 5{\it b} covering the frequency range from 0.09
to 5.2 mHz. Figure 5{\it e} is a spline-smoothed fit to the lower
envelope of the power spectra to show the enhanced background power toward the
limbs.

\begin{figure}[!h]
\setcounter{figure}{5}
\includegraphics[width=3.0in,keepaspectratio=true,clip=true]{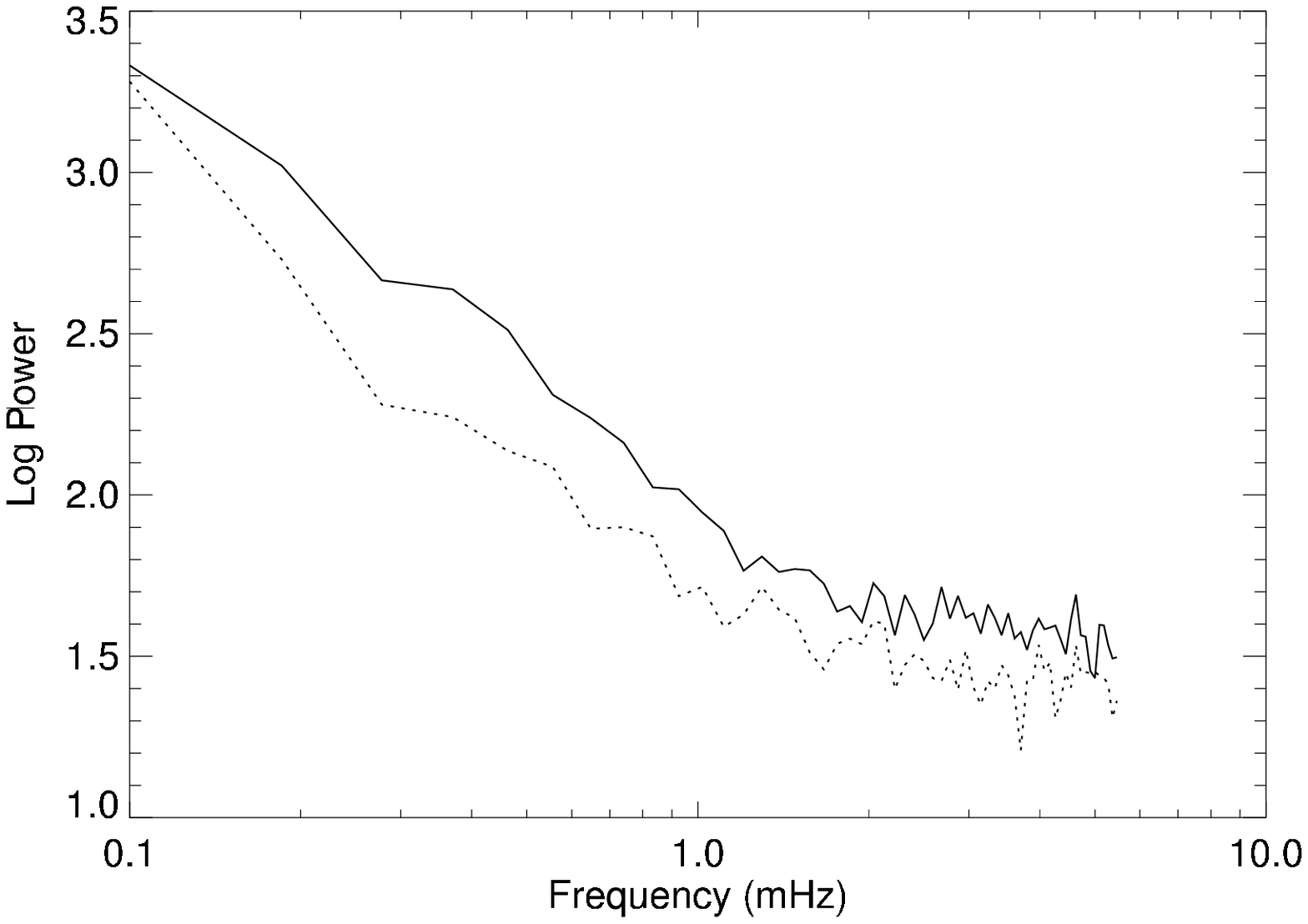}
\caption{Typical power spectrum averages for a
region near the limb ({\it solid line}) and near the disk center ({\it
dotted line}). See text for discussion.}
\end{figure}

Going beyond a simple rms analysis, Figure 6 shows average power
spectra for data near the limb (full line) and near disk center (dotted
line). The near-limb spectrum is dominated by a $\nu^{-1.4}$ slope at low
frequency up to about 1.5 mHz. At higher frequencies, after
a short transition, the spectrum is essentially flat due to instrumental and
registration noise. There is no obvious indication of excess power around
3 mHz. Near the disk center, where the horizontal
field component is small, the average power spectrum is weaker and more
complicated. At low frequencies it becomes steeper than -1.4. 
From 0.3 to $\sim$2.0 mHz the power varies as $\nu^{-1}$.
At higher frequencies the spectrum is flattened by noise with no sign
of extra power at 3 mHz.

\section{Summary and Discussion}
We discovered a ubiquitous, nearly horizontal component of the solar magnetic
field in quiet regions of the photosphere. Its reality is confirmed using
observations with different instruments, spectrum lines and measurement
techniques. This component exhibits wide ranges of spatial and temporal
scales: from our resolution limit of a few arcsec up to $\sim$15\arcsec\
and from times of several minutes to hours. In movies of the LOS
field, this component looks like a seething pattern of mottling.
At a spatial resolution of 5\arcsec\, the average temporal rms of the
horizontal field variation near the limb is 1.4 G. Doubling the spatial
resolution to 2\farcs5 increases the temporal rms value to 1.7 G.

Based on its temporal and spatial scales, we speculate that the seething
horizontal field is driven by granular and supergranular convection,
and by field line reconfigurations in response to evolving flux distributions
in the nearby network and IN. \citet{dep02} observed eruption and subsequent
shredding of a few IN magnetic flux elements consistent with this notion.
If connections with existing network flux elements are important, we might 
expect a dependence of the strength of the horizontal component upon the
amount of neighboring large-scale magnetic flux. The lower right part of
Figure 3 contains such a location and there is no evidence to support this
expectation. This finding, and the absence of any evidence of latitudinal or
longitudinal dependence, favors the idea that the horizontal field is mainly
created and driven by local processes. Recent numerical
simulations by \citet{geo07} show that it is impossible to separate
temporal and spatial components of a solar convective turbulence spectrum.
So we cannot make a simple interpretation of the observed
$\nu^{-1.4}$ power variation over more than a decade of frequency.
The absence of any but feeble hints of excess power at 3 mHz
suggests that p-mode oscillations play little, if any, role in the
dynamics of the horizontal field.

The sporadic HIFs observed by \citet{lit96} are probably small bipolar
flux elements erupting from the interior. They are too infrequent to
explain the ubiquitous, nearly horizontal field fluctuations that we found.
Numerical magnetoconvection simulations can provide insight into the physics
of the horizontal field. For example, \citet{gad01} find a weak, predominantly
horizontal field in the photospheric layers of granules associated with
strong horizontal flows. It would be interesting to see what more
advanced models, such as those of \citet{kho05}, would predict
on a scale larger than granulation. More extensive and detailed 
observations are certainly required to clarify the physical nature of
the nearly horizontal field. Many questions remain unanswered: Is a
similar field observed higher in the solar atmosphere? Is the field a
miniature version of canopy fields found around strong flux
concentrations. What is the detailed
relationship with the photospheric convective and oscillatory velocity 
fields? What is the association with the granular magnetic field of
\citet{lin99}? How does the field fit with Hanle-effect
observations of a microturbulent magnetic field (e.g. Stenflo, Keller,
\& Gandorfer 1998)?

Although most attention has been directed to the vertical component of
the quiet Sun magnetic field, the ubiquitous presence of a nearly
horizontal component suggests that additional studies may prove it to
be at least as significant in improving our understanding of
solar magnetism. A practical consequence of the field involves
extrapolations of photospheric field measurements to the corona.
It is usually assumed that the observed LOS fields are radially oriented
in order to estimate the surface distribution of magnetic flux. This
assumption is acknowledged as wrong in active regions. Now we know that it is
also incorrect for quiet regions observed near the limb.
However, time and spatial averaging of observations may 
mitigate this effect. Finally, we note that it is overly simplistic to
consider the IN field as being composed of independent vertical and
horizontal components. It is most likely that these are
just observational manifestations of a dynamically interacting field of
wonderful complexity.

\acknowledgments

We gratefully acknowledge the SOLIS and GONG team members for their
devoted and skilled work that provides high-quality data to the
research community. SOLIS/VSM data used here were produced with
support from NSF and NASA. This work utilizes data obtained by
the Global Oscillation Network Group (GONG) Program, managed
by the National Solar Observatory. The real-time GONG data were acquired by
instruments operated by the Big Bear Solar Observatory and Cerro Tololo
Interamerican Observatory and obtained from publically available archives.
The National Solar Observatory is operated by the Association of
Universities for Research in Astronomy, Inc. (AURA), under
cooperative agreement with the National Science Foundation.

{\it Facilities:} \facility{GONG}, \facility{SOLIS (VSM)}.

\onecolumn
\begin{figure}[!h]
\setcounter{figure}{4}
\includegraphics[width=7.0in,keepaspectratio=true,clip=true]{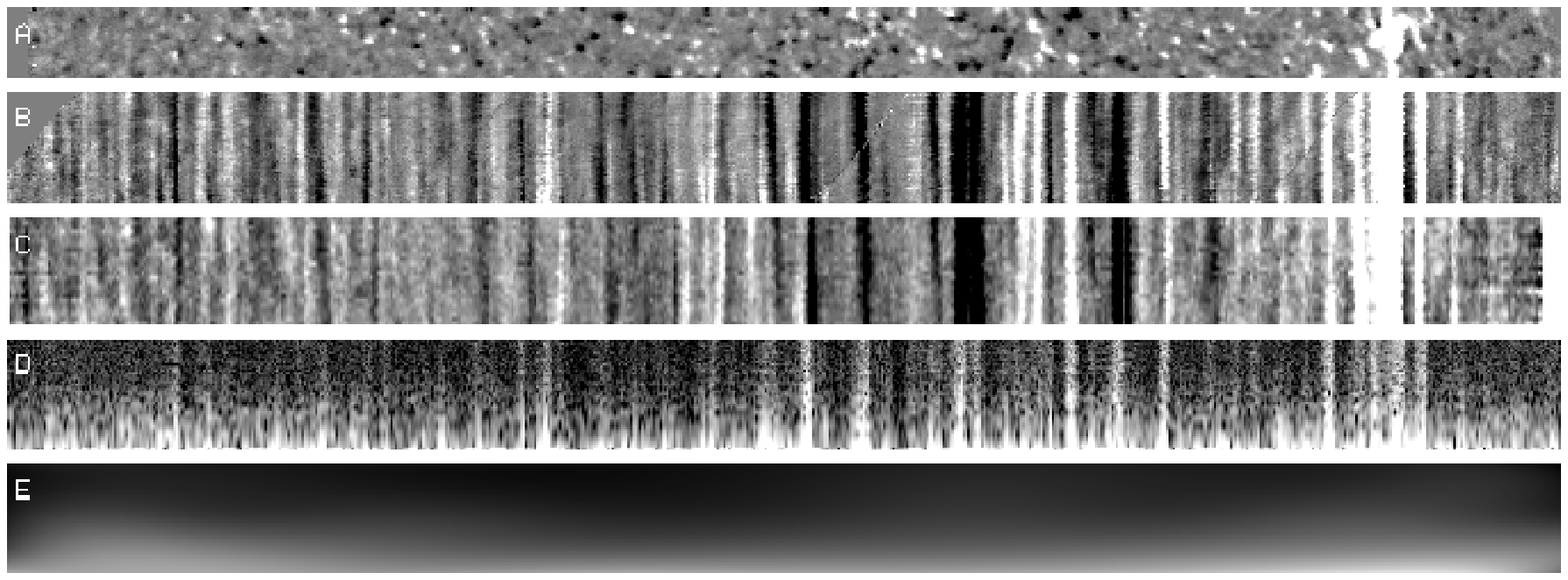}
\caption{Limb-to-limb cuts across the solar disk on 2006 December 15.
({\it a}) A VSM LOS magnetogram displayed with saturation at $\pm$22 G.
({\it b}) VSM data time variation for 3.2 h along a trace
through the area shown in ({\it a}). The display saturates at $\pm$11 G
and the spatial glitch in the middle is a data reduction artifact.
({\it c}) Same except using GONG data. Registration with ({\it b}) is
best near the limbs. Coarser spatial and time GONG samples were
interpolated to match the VSM data. Note higher noise level compared
to VSM data. ({\it d}) Log of 3 decades of power spectra of the columns
in ({\it b})
displayed on a log frequency scale from 0.09 at the bottom to 5.2 mHz
at the top. Note absence of any obvious periodic signal.
({\it e}) Spline-smoothed fit of the background of ({\it d}).
Note the larger background power levels toward the limbs.}
\twocolumn
\end{figure}

\end{document}